\documentstyle[aps,prb,epsfig,floats]{revtex}

\begin{document}
\twocolumn

\title{Dynamics of Excited Electrons in Copper: Role of Auger Electrons}
\author{R.~Knorren$^*$, G.~Bouzerar, and K.~H.~Bennemann}
\address{Institut f\"ur Theoretische Physik, Freie Universit\"at
Berlin, Arnimallee 14, D-14195 Berlin, Germany} 
\date{\today}  
\maketitle

\begin{abstract}
Within a theoretical model based on the Boltzmann equation,  we
analyze in detail the structure of the unusual peak recently observed
in the relaxation time in Cu. In particular, we discuss the role of
Auger electrons in the electron dynamics and its dependence on  the
$d$-hole lifetime, the optical transition matrix  elements and the
laser pulse duration.  We find that the Auger contribution to the
distribution is very sensitive to both the $d$-hole lifetime $\tau_h$
and the laser pulse duration $\tau_l$ and can be expressed as a
monotonic function of $\tau_l/\tau_h$. We have found that for a given
$\tau_h$, the Auger contribution is significantly smaller for a short
pulse duration than for a longer one.  We show that the relaxation
time at the peak depends linearly  on the $d$-hole lifetime, but
interestingly not on the amount of Auger electrons generated. We
provide a simple expression for the relaxation time of excited
electrons which shows that its shape can be understood by a phase
space argument and its amplitude is governed by the  $d$-hole
lifetime.  We also find that the height of the peak depends on both
the ratio of the optical transition matrix elements  $R=|M_{d \to
sp}|^2/|M_{sp \to sp}|^2$   and the laser pulse  duration. Assuming a
reasonable value for the ratio, namely  $R = 2$, and a $d$-hole
lifetime of $\tau_h=35{\rm\ fs}$, we obtain for the calculated height
of the peak $\Delta \tau_{th}=14$~fs, in  fair agreement with $\Delta
\tau_{exp} \approx 17$~fs measured for polycrystalline Cu.
\end{abstract}

\section{Introduction}
Two-photon photoemission (2PPE) has been applied intensively to study
the electron-electron interaction in  alkali, noble and transition
metals.\cite{Schmuttenmaer94,Hertel96,Aeschlimann96,Ogawa97a,Aeschlimann97}
The experimental results are often compared with the  Fermi-liquid
theory (FLT),\cite{Quinn62} but such a comparison is only valid if the
influence of secondary-electron generation and transport of  excited
carriers is negligible. Indeed, a series of 2PPE experiments involving
photoexcitation of electrons from  the $d$ band in Cu have shown an
unusual non-monotonic behavior of the relaxation
time.\cite{Pawlik97,Cao97,Knoesel98,Petek99}  The relaxation time
shows a broad peak at energies just above the $d$-band threshold which
shifts linearly with photon energy. Both the non-monotonic behavior
and the dependence on photon energy are incompatible with FLT
description  or with recent calculations based on  Green's-function
formalism.\cite{Campillo99,Echenique00,Schoene99,Keyling00}  There is
currently a controversy in the literature on the origin of the peak
and especially on the role of secondary
electrons.\cite{Pawlik97,Cao97,Knoesel98,Petek99}   One of the key
questions concerns the amount of Auger electrons contributing to the
hot-electron distribution in the region of the peak. On the basis of
simple model calculations,  Petek {\it et al.} have
estimated that the Auger contribution is less than 10\%,\cite{Petek99b} while
Knoesel, Hotzel, and Wolf could reproduce their
correlation  measurement by assuming that the hot electron population
above the $d$-band threshold was entirely generated via the Auger
process.\cite{Knoesel98}  
Note that the experiments were performed using different 
pulse durations (12 and 70~fs, resp.). 

In a previous work, we have developed a theory for the dynamics of
excited electrons in metals, which includes both the effects of
secondary electrons and transport. We used a Boltzmann-type  equation
in the random-{\bf k} approximation  for the calculation of the distribution
function of excited electrons. A detailed description is given in
Ref.~\onlinecite{Knorren99b}. The calculations have shown a peak in
the relaxation time at the right energy reflecting the $d$-band
threshold and a linear shift with photon frequency, in agreement with
experiments. However, the height of the peak $\Delta \tau_{th} \approx
10$~fs calculated using $\tau_h=9$~fs was smaller than the observed
one $\Delta \tau_{exp} \approx 17$~fs.\cite{Cao97} Furthermore, a
simplified procedure was used to extract the relaxation time which
also led to an overestimate of the peak height. Indeed for simplicity,
the relaxation time was extracted by fitting the distribution of
excited electrons, $f(t)$, which was very
sensitive to the influence of secondary electrons causing a deviation
from a simple exponential decay. To make our results more reliable,
in this work we use the same fitting procedure as the one used by
experimentalists, i.e. we fit  $I^{\rm 2PPE}(\Delta t)$, given by the
convolution of the probe laser profile with the distribution of
excited electrons. 

In this paper, we  will analyze in detail the structure and height of
the peak and the role of secondary  electrons, especially focussing on
the Auger contribution and the $d$-hole lifetime.  In order to shed
light on the  contribution of Auger
electrons,\cite{Knoesel98,Petek99b} we have started by analyzing their
weight in the total hot-electron distribution. For this purpose we
have performed calculations for different hole lifetimes ($\tau_h$)
and laser pulse durations ($\tau_l$) and found that their relative
contribution can be expressed in a simple way as a product of a
function of $\tau_l/\tau_h$ and a phase space function which is
energy-dependent.  In addition, we show that the height of the peak in
the relaxation time depends  mainly on the hole lifetime and
surprisingly not on the amount of Auger electrons. It will be shown
that the relaxation time for energies  above the $d$-band threshold
scales linearly with $\tau_h$ and its structure is determined by the
same phase space function as mentioned above.  Finally, we study how
the height of the peak depends on the optical transition matrix
elements and on the laser  pulse duration.

\section{Results and discussion}
\begin{figure}[b]
\begin{center}
\epsfig{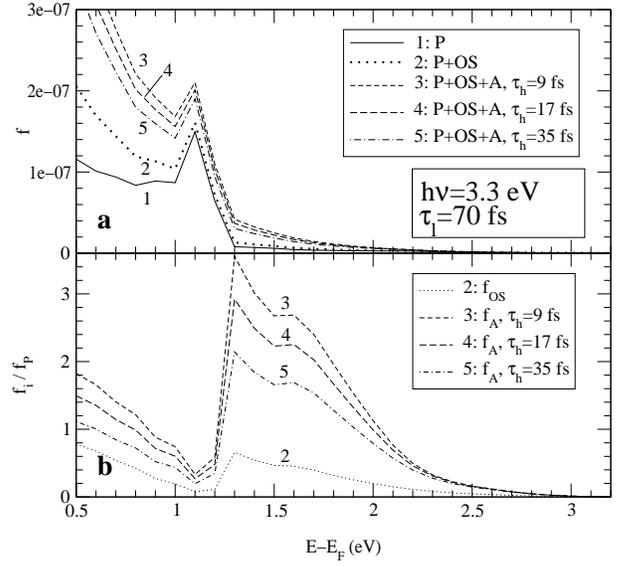} 
\end{center}
\caption{\label{fig1} a) Distribution $f(E)$ of hot electrons at t=0 
  for different cases (1: only primary, 2: primary and other
  secondary, 3--5: primary, other secondary and Auger electrons for
  different values of the hole lifetime $\tau_h$).
  b) Ratio of distribution of Auger ($f_A$) and other secondary
  electrons ($f_{OS}$) to
  distribution of primary electrons ($f_P$). The pulse duration is
  $\tau_l=70$~fs and   the photon energy $h\nu=3.3$~eV.}
\vspace*{-0cm}
\end{figure}

As said in the introduction, the composition of the hot-electron
distribution in terms of different contributions ($f(E)=\sum_i f_i(E),
i=P, A, OS$ for primary, Auger, other secondary electrons) is an open
question. We determine their respective weight and compare the
distribution of hot electrons for different cases in Fig.~\ref{fig1}.
We especially focus our attention on the  variation as function of the
hole lifetime at the top of the $d$ band, $\tau_h$. The hole lifetime
plays an important role, since it determines the generation rate of
Auger electrons which may strongly influence the dynamics of excited
electrons above the Fermi energy. Note that in our model, the hole
lifetime is the  inverse of the Auger scattering rate, thus we cannot
vary  independently the hole lifetime and the amount of Auger
electrons generated.  One should mention that experimentally, lower
bounds of 24~fs and 26~fs have been reported for the  hole lifetime at
the top of the $d$ band in Cu.\cite{Petek99a,Matzdorf99} Since there
is up to now no clear consensus on the value of $\tau_h$, in our
calculation we will consider it as a free parameter. In absence of a
$d$ band, the electron-hole symmetry would hold, and our calculation
gives then $\tau_h = 9$~fs for the hole lifetime at
$E-E_F=-2.2$~eV. Due to the localized nature of the $d$-wave
functions, the $d$-hole lifetime is expected to be significantly
larger than the $s$-hole lifetime, which will hence be considered as a
reasonable lower bound.  We start with the case of a laser pulse of
photon energy $h\nu=3.3$~eV and duration $\tau_l=70$~fs (FWHM) as used
in Ref.~\onlinecite{Knoesel98}. Later, we will discuss the effect of
changing the laser pulse duration.

In Fig.~\ref{fig1}(a) we show results of the distribution of hot
electrons for different cases: with and without secondary electrons
and for different hole lifetimes. We observe a peak in the primary
electron distribution at $E-E_F=1.1$~eV which is due to the sharp
feature in the $d$-band density of states; at this energy, primary
electrons are dominating.   To simplify the discussion, in
Fig.~\ref{fig1}(b), we plot $f_i/f_P$ for $i=A, OS $.  In the region
of the peak in $f$, the  primary electron contribution is dominating,
which appears now as a dip in $f_i/f_P$.  Below the peak, there is a
continous rise of both the Auger and other secondary electron
contribution. On the other hand,  just above the peak, there is a
sharp increase of  the weight of the secondary electrons, largely
dominated by the Auger contribution, with   a maximum at $E-E_F =
1.3$~eV. We observe that a change of the hole lifetime leads to a
significant variation of the Auger contribution.  For example, for
$\tau_h=35$~fs, $f_A/f_P=2.2$, whereas for shorter lifetime
$\tau_h=9$~fs, $f_A/f_P=3.6$. In our model, a shorter hole lifetime
corresponds to a larger Auger scattering rate and therefore to a
larger amount of Auger electrons. Note that $f_A/f_P$ has the same
shape for any given $\tau_h$, only the relative amplitude is
changing. This will be analyzed further when discussing the effect of
the laser pulse duration.  Furthermore, we find that for
$E-E_F>2.5$~eV, the contribution of secondary electrons is always
negligible.

\begin{figure}[htb]
\vspace*{0cm}
\begin{center}
\epsfig{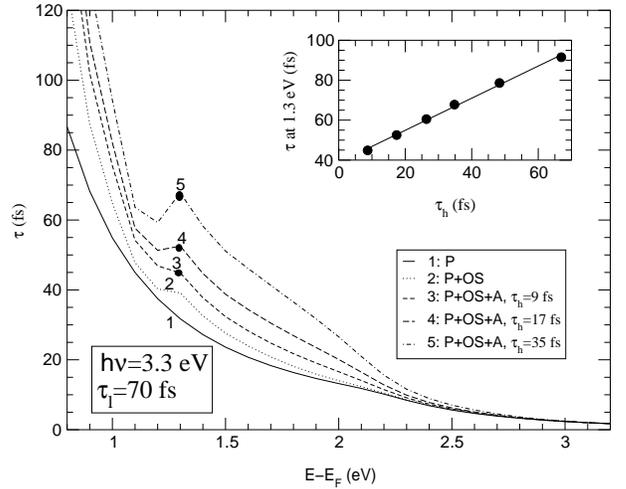} 
\end{center}
\caption{\label{fig2} Dependence of excited-electron 
relaxation time (determined by a fit to the 2PPE signal)
on hole lifetime $\tau_h$ as a function of energy. The inset shows
the relaxation time at the peak 
position, $E-E_F=1.3$~eV, as a function of $\tau_h$.}
\end{figure}

In Fig.~\ref{fig2}, we show the relaxation time as function of energy
for different cases (see Fig.~\ref{fig1}). The relaxation time is
determined by a fit to the 2PPE signal as discussed in connection with
Fig.~\ref{fig6}.  It will be interesting to see how the peak in the
relaxation time  depends on both $f_A/f_P$ and on the hole lifetime.
First, we observe that when considering only primary processes
($\tau(E)=\tau_P(E)$), the relaxation time is a monontonic function,
as it can be expected on the basis of FLT. However, when including
other secondary  processes (but no Auger process), we immediately
observe some significant changes for $E-E_F<2.3$~eV and some new
structure at intermediate energy.  When comparing with
Fig.~\ref{fig1}(b), the changes in $\tau(E)$ due to the presence of
other secondary electrons coincide with  the variation of $f_{OS}/f_P$
discussed previously. Indeed, at $E-E_F=1.1$~eV (dip in $f_{OS}/f_P$),
$\tau$ is almost unchanged. Then a small peak appears at
$E-E_F=1.3$~eV (region where the secondary contribution is the
highest), and for $E-E_F>2.5$~eV, $\tau(E)=\tau_P(E)$ (negligible
secondary electron contribution). This picture remains qualitatively
unchanged when including the Auger process.

Let us now comment on the quantitative aspect. First of all, a clear
and well-defined  peak is observable when $\tau_h$ gets sufficiently
large (typically $\tau_h>17$~fs).   It is also interesting to note
that for small $\tau_h$ (large Auger scattering rate), the effect is
very weak, in spite of the fact that the Auger contribution to the
electron distribution is large [see Fig.~\ref{fig1}(b)]. This is a
clear indication that the amplitude of the peak in $\tau$ is mainly
controlled by the hole lifetime. This is indeed illustrated in the
inset of Fig.~\ref{fig2}, where $\tau$ at $E-E_F=1.3{\rm\ eV}$ is
plotted as function of the hole lifetime. We see clearly that $\tau$
scales linearly with $\tau_h$.

It is important to mention that we have found that  this linear
scaling behavior is also valid for other energies. This suggests that
$\tau(E)$ could be expressed as a sum of two terms, $\tau_0(E)$
(including primary and other secondary processes), and a term
proportional to $\tau_h$ (Auger contribution), where the coefficient
is a phase space  factor depending on the DOS only. To be more
explicit, we suggest that the relaxation time can be written as
$\tau(E)-\tau_0(E) \propto \tau_h\ F(E)$, where $F(E)$ is a phase
space factor defined as the ratio between a) the available phase space
for  Auger scattering and b) the phase space for the optical
excitation.

\begin{figure}[htb]
\vspace*{0cm}
\begin{center}
\epsfig{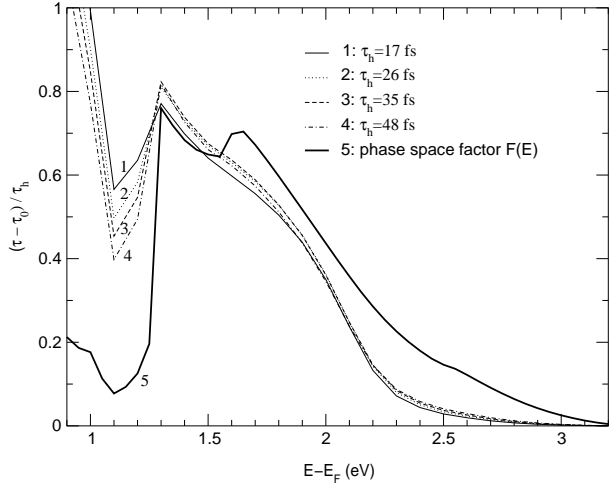} 
\end{center}
\caption{\label{fig3} Ratio $[\tau(E) - \tau_0(E)]/\tau_h$ discussed in
  the text, where $\tau_0(E)$ includes only effects of primary and other
  secondary electrons (curve 2 in Fig.~\ref{fig2}). $F(E)$ is the
  phase space factor discussed in the text.}
\end{figure}

To test the validity of this expression, in Fig.~\ref{fig3}, we  have
plotted in the same figure  $[\tau(E)-\tau_0(E)]/\tau_h$ as  function
of $E-E_F$ and the calculated $F(E)$.  Clearly, for  $E-E_F>1.3$~eV,
one can see that the values for $[\tau(E)-\tau_0(E)]/\tau_h$ lie
almost on the same curve, which depends only
on the energy. A deviation is observed for smaller energy, where Auger
electrons are no longer the dominant contribution to the hot-electron
distribution [see Fig.~\ref{fig1}(b)].  However, the position of the dip
is always found at 1.1~eV. Within some deviations at high and low
energy, we observe a fair agreement between $F(E)$ and
$[\tau(E)-\tau_0(E)]/\tau_h$.  Note that a peak appears in $F(E)$ at
$E-E_F \approx 1.7$~eV, whereas in $[\tau(E)-\tau_0(E)]/\tau_h$, one
finds only a shoulder at this position. This can be attributed to the
fact that by definition $F(E)$ contains the explicit details of the
DOS, whereas the  relaxation time is not so sensitive to the details
of the DOS. 

To summarize this discussion, the important result is that the
amplitude of the broad peak in $\tau(E)$ is controlled by $\tau_h$ and
its shape by the phase-space factor $F(E)$ (in random-{\bf k}
approximation, it depends only on the DOS). When comparing
Fig.~\ref{fig3} and Fig.~\ref{fig1}, we find that $f_A/f_P$ is also
proportional to $F(E)$, which will be commented later.

\begin{figure}[htb]
\begin{center}
\epsfig{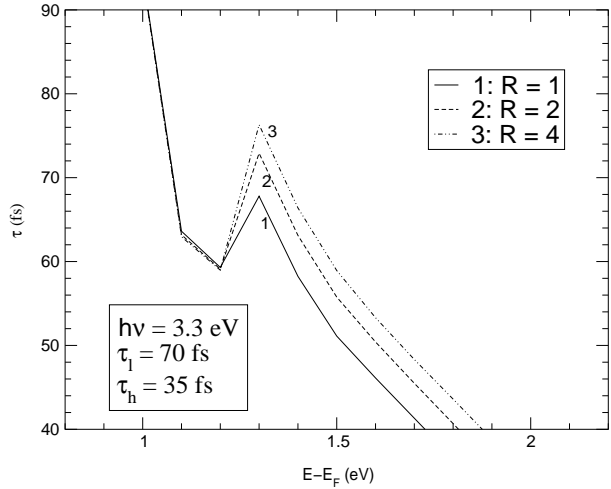} 
\end{center}
\caption{\label{fig4} Dependence of the relaxation time on the ratio
of optical transition matrix elements $R=|M_{d \to
sp}|^2/|M_{sp \to sp}|^2$. The laser pulse duration is $\tau_l=70$~fs and
the hole lifetime is  $\tau_h=35$~fs.}
\end{figure}

As was shown in Fig.~\ref{fig3}, the shape of the peak in the
relaxation time is determined by the function $F(E)$ which contains
the information on the optical transition matrix elements. Therefore,
in Fig.~\ref{fig4} we analyze how sensitively the relaxation time
$\tau(E)$ depends on the optical transition matrix elements.  Till
now, we have presented data for the case of equal matrix elements,
$M_{d \to sp}=M_{sp \to sp}$.    However, on the basis of the
different degree of localization of the $d$ and $sp$-type wave
functions,  it is expected that $M_{d \to sp}$ should be larger than
$M_{sp \to sp}$. For example, in the calculation of the dielectric
function for Ag, a ratio  $|M_{d \to p}|^2/|M_{p \to s}|^2=2.21$ was
estimated.\cite{Rosei74}  

In Fig.~\ref{fig4}, we have plotted  $\tau(E)$ for different values of
the ratio $R=|M_{d \to sp}|^2/|M_{sp \to sp}|^2$.  If we define the
height of the peak as $\Delta \tau = \tau(1.3{\rm\ eV}) -
\tau(1.2{\rm\ eV})$, we find that when changing $R$ from 1 to 2,
$\Delta \tau$ varies significantly from 9~fs to 14~fs (increase by
60\%). However for a further increase of $R$ from 2 to 4, the change
is much weaker (14~fs to 17~fs, increase by only 30\%).  Assuming that
for Cu, $R$ is of the same order of magnitude as for Ag, one can get
an estimate of the hole lifetime required to get fair agreement with
the observed height of the peak for polycrystalline Cu, $\Delta
\tau_{exp}\approx 17$~fs.\cite{Cao97} We  have found that for $R=2$,
$\tau_h=35$~fs gives $\Delta \tau_{th}=14$~fs. Note that our results
obtained in the random-{\bf k} approximation are most suitable for
comparison with experiments performed on polycrystalline
material. Measurements on single crystals for different orientations
have provided values ranging from $\Delta \tau_{exp}=15$ to 40~fs,
which is the same order of
magnitude.\cite{Pawlik97,Cao97,Knoesel98,Petek99}  It is interesting
to note that $\tau_h=35$~fs coincides with the value used by Knoesel,
Hotzel, and Wolf in their simulation.\cite{Knoesel98}

Note that our conclusions are not restricted to the case of Cu only,
but hold also in the case of other noble and transition
metals. Indeed, let us discuss the case of Ag, Au and the 3$d$
transition metals Fe, Co and Ni. In the case of Ag, no peak is
observed in the relaxation time,\cite{Aeschlimann96} due to the fact
that the $d$ band threshold is  approximatively at 4~eV below the
Fermi level and no $d$ electrons are excited for the photon energies
widely used in 2PPE experiments  ($h\nu < 4$~eV). The case of Au is
more intriguing, since the location of the $d$-band threshold is very
similar to the one in Cu. Indeed, a very weak structure in the
relaxation time was also observed  for Au.\cite{Cao98} The fact that
the peak in Au is so weak can be attributed to the small degree of
localization of the of the $d$ hole.  It is expected that a  5$d$ hole
in Au should be less localized and therefore have a larger Auger
scattering rate than a 3$d$ hole in Cu. Thus one expects $\tau_h^{\rm
Au} < \tau_h^{\rm Cu} $ and thus in accordance with our result that
the peak is governed by $\tau_h$, a less pronounced structure in Au
than in Cu. Another possible explanation is the fact that the peak in
the $d$-band DOS in  Au is much less pronounced than in Cu (again due
to the less localized $d$ electrons in Au), leading to a less sharp
separation between excitations from the $d$ and $sp$ bands and
therefore to a weaker peak.  Let us now comment on the fact that in
the ferromagnetic transition metals Fe, Co and Ni, no peak is
observed.\cite{Aeschlimann97,Knorren99b} There are  two main reasons
for the absence of this feature. (i) A threshold for $d \to sp$
transitions is not observable due to the band structure.  The majority
$d$ band extends to energies above $E_F$ and a  $d \to sp$ threshold
does not exist. The upper minority $d$ band edge is located very
closely to $E_F$ and the threshold occurs only at the highest
excitation energies where secondary (Auger) electron contributions are
not important.  (ii) The $d$ holes have a large available phase space
for decay within the $d$ band which leads to a very short $d$-hole
lifetime.

\begin{figure}[htb]
\vspace*{0cm}
\begin{center}
\epsfig{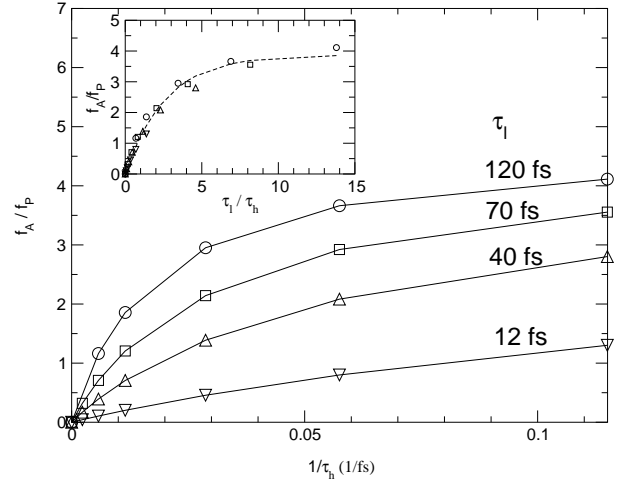} 
\end{center}
\caption{\label{fig5} Ratio of Auger to primary electron
  contribution, $f_A/f_P$ at $E-E_F=1.3$~eV, as a function of the inverse
  hole lifetime $1/\tau_h$
  for different laser pulse durations $\tau_l$. In the inset, we show
  $f_A/f_P$ as a function of $\tau_l/\tau_h$. The line is a guide to
  the eye.}
\end{figure}

\begin{figure}[htb]
\vspace*{0cm}
\begin{center}
\epsfig{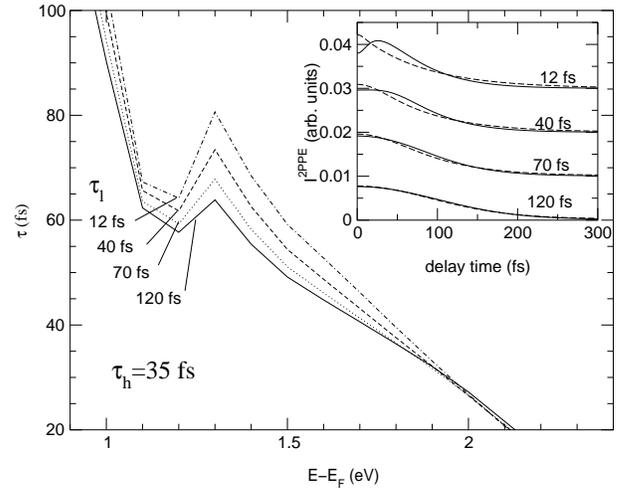} 
\end{center}
\caption{\label{fig6} Relaxation time for different laser pulse
  durations $\tau_l$. The hole lifetime used is $\tau_h=35$~fs. 
  In the inset, we show the calculated two-photon photoemission
  intensity $I^{\rm 2PPE}$ at $E-E_F=1.3$~eV (solid line) together
  with the fit assuming exponential 
  decay from which the relaxation time is determined (dashed line).} 
\end{figure}

All the results presented up to now were for
$\tau_l=70$~fs. However, experiments were performed with laser pulse
durations ranging from  70~fs down to
12~fs.\cite{Pawlik97,Cao97,Knoesel98,Petek99} Hence, it is
interesting to calculate how $\tau(E)$ depends on $\tau_l$.  First,
let us show that the relative weight of the Auger contribution to the
hot electron distribution is indeed strongly sensitive to the laser
pulse duration. 
In Fig.~\ref{fig5}, we have plotted $f_A/f_P$ as a
function of $1/\tau_h$ for laser pulses varying from 12 to 120~fs
duration. The first observation is that when reducing the laser pulse
duration for a given $\tau_h$, $f_A/f_P$ is strongly reduced. For
instance, for $\tau_h=35$~fs, $f_A/f_P$ goes from 3 to 0.5 (6 times
smaller) when reducing the laser pulse from 120 to 12~fs.  This might
already explain the origin of the controversy about the amount of the
Auger electron contribution mentioned before. It is clear that in
experiments with very short laser pulses (12~fs in
Ref.~\onlinecite{Petek99b}), the Auger electron contribution is much
smaller than in experiments with longer laser pulses (70~fs in
Ref.~\onlinecite{Knoesel98}).  At this level, one can conclude that
the laser pulse duration $\tau_{\rm l}$ is a new relevant time
scale. However, we have found that the only relevant parameter for the
variation of $f_A/f_P$ is in fact $\tau_{\rm l}/\tau_h$. In the inset
of Fig.~\ref{fig5}, we have plotted $f_A/f_P$ as function of
$\tau_{\rm l}/\tau_h$ and we see that all the data points lie to a
very good approximation on the same curve, i.e. $f_A/f_P(\tau_{\rm l},
\tau_h) \propto g(\tau_{\rm l}/\tau_h)$. In this figure, the data are
plotted for $E-E_F=1.3$~eV, however, it has to be stressed that this
scaling is also found for higher energies. The ratio $f_A/f_P$ can
be reduced to an expression of the form  $f_A/f_P(E, \tau_{\rm l},
\tau_h) \propto g(\tau_{\rm l}/\tau_h) F(E)$, where  $F(E)$ is again
the phase factor discussed in the previous section.

Since the relaxation time is determined from $I^{\rm 2PPE}(\Delta t)$
which contains explicitly the information about the laser pulse, let
us now analyze how $\tau$ is affected by changing the pulse duration.
In Fig.~\ref{fig6}, we show the dependence of the relaxation time
$\tau(E)$ on the pulse duration for durations varying from 120 to
12~fs. We observe a significant effect in the region of the peak.
Interestingly, the peak gets higher for the shorter pulses, although
the amount of Auger electrons is much smaller, see
Fig.~\ref{fig5}. Note that the dependence on the pulse duration
disappears for high energies, where the secondary-electron
contribution is negligible.  In order to understand the sensitivity to
the pulse duration, in the inset of Fig.~\ref{fig6}, we present the
calculated $I^{\rm 2PPE}$ and the fit by an exponential model function
used to extract the relaxation time. We see that the agreement is very
good for pulses of 120 and 70~fs duration, while there is a clear
deviation from the single-exponential fit for shorter pulses of 40 and
12~fs.  Indeed, for the 12~fs pulse, there is a delayed rise in
$I^{\rm 2PPE}$ which is due to the delayed generation of Auger
electrons. Such a delayed rise for very short laser pulses was
recently observed experimentally.\cite{Petek99} As an additional
remark, the observed deviations underline the difficulties in
extracting a relaxation time in the limit of very short pulses and
indicate that maybe an improved definition of the relaxation time
would be appropriate in this case.

\section{Conclusion}
To conclude, this work presents a detailed analysis of the role of
secondary electrons in 2PPE experiments on Cu.  The $d$-hole lifetime
plays a crucial role.  Our conclusions are more general and can also
be applied to other noble and transition metals as discussed. We have
found that in the case of Cu, the secondary-electron distribution is
dominated by the Auger contribution (for $\tau_h\approx 35$~fs) for
longer laser pulses ($\tau_l>40$~fs), but not for a very short laser
pulse of $\tau_l=12$~fs. For a given $\tau_h$, the Auger contribution
is much larger for a longer laser pulse duration than for a shorter
one.  The parameter which controls the variation of $f_A/f_P$ is
$\tau_l/\tau_h$. Concerning the structure in the relaxation time, we
have shown that the relaxation time at the peak depends linearly on
$\tau_h$, but surprisingly not on the amount of secondary electrons
generated. The shape depends on a phase-space factor (in random-{\bf k}
approximation, on the DOS). We provided an expression for the
relaxation time $\tau(E)$  as a sum of two terms: the first for
primary (and other secondary) electrons and the second for the Auger
contribution.  We have also found that the height of the peak
depends sensitively on the optical transition matrix element ratio,
$R=|M_{d \to sp}|^2/|M_{sp \to sp}|^2$, and on the laser pulse
duration. For a value of $R = 2$ and $\tau_h=35{\rm\ fs}$, the
calculated height of the peak is $\Delta \tau_{th}=14$~fs in fair
agreement with a measurement on polycrystalline Cu giving $\Delta
\tau_{exp}\approx 17$~fs.
Note, preliminary results indicate that transport changes
the magnitude of the relaxation time, but not the structure and height
of the peak. The influence of transport will be studied in detail in a
separate publication.\cite{Knorren00} 
In view of the importance of the $d$-hole
lifetime, it would be highly desirable to perform further experiments
as well as theoretical calculations on the $d$-hole lifetime in both
Cu and Au.

\acknowledgments
We would like to thank M.~Aeschlimann, M.~Wolf and E.~Matthias
for interesting discussions. Financial support by Deutsche
Forschungsgemeinschaft, Sfb 290 and 450, is gratefully acknowledged.

\end{document}